\documentclass[fleqn,usenatbib]{mnras}

\usepackage{newtxtext,newtxmath}

\usepackage[T1]{fontenc}
\usepackage{ae,aecompl}

\usepackage{siunitx}
\usepackage{placeins}

\usepackage{graphicx}	
\usepackage{amsmath}	
\usepackage[dvipsnames]{xcolor}

\newcommand{\halpha}{H$\alpha$}
\newcommand{\novacar}{ASASSN-18fv}
\newcommand{\novasgr}{Nova V5668 Sgr}
\newcommand{\mgab}{YZ Reticuli}
\newcommand{\pnvj}{V6568 Sgr}
\newcommand{\RSoph}{RS Oph}
\newcommand{\ssf}{SS433}
\newcommand{\DQHer}{DQ Her}
\newcommand{\fermilat}{Fermi-LAT}
\newcommand{\asassn}{ASAS-SN}
\newcommand{\nustar}{NuSTAR}
\newcommand{\vlba}{VLBA}
\newcommand{\Aquila}{Aquila}
\newcommand{\codeendeavour}{{\tt endeavour}}
\newcommand{\codepoirot}{{\tt poirot}}
\newcommand{\Hires}{LHires}
\newcommand{\gaussian}{Gaussian}
\newcommand{\gaussians}{Gaussians}

\newcommand{\about}{$\sim\,$}
\newcommand{\GJW}{Global Jet Watch}

\definecolor{orange2}{RGB}{226, 117, 32}

\newcommand{\kms}[1]{$#1\,\si{\km\per\second}$}
\newcommand{\cn}{classical nova}
\newcommand{\rn}{recurrent nova}
\newcommand{\rne}{recurrent novae}
\newcommand{\cne}{classical novae}

\newcommand{\Cn}{Classical nova}

\newcommand{\Cne}{Classical novae}

\newcommand{\angstrom}[1]{$#1\,\si{\angstrom}$}
\newcommand{\days}[1]{$+#1\,$d}
\DeclareSIUnit \parsec {pc}
\newlength{\wdth}

\title[Jets in classical nova YZ Reticuli]{The precessing jets of classical nova YZ Reticuli}

\author[D. McLoughlin et al.]{
Dominic McLoughlin,$^{1}$\thanks{E-mail: dominic.mcloughlin@physics.ox.ac.uk (DM)}
Katherine M. Blundell,$^{1}$
Steven Lee, $^{2, 3}$
Chris McCowage $^{2, 3}$
\\
$^{1}$Department of Physics, University of Oxford, Keble Rd, Oxford OX1 3RH, United Kingdom\\
$^{2}$Research School of Astronomy and Astrophysics, Australian National University, Canberra, ACT 2611\\
$^{3}$Anglo-Australian Telescope, Coonabarabran NSW 2357, Australia\\
}

\date{Accepted 2021 February 23. Received 2021 February 23; in original form 2021 January 08}

\pubyear{2020}

\begin{document}
\label{firstpage}
\pagerange{\pageref{firstpage}--\pageref{lastpage}}
\maketitle

\begin{abstract}
The classical nova YZ Reticuli was discovered in July 2020. Shortly after this we commenced a sustained, highly time-sampled coverage of its subsequent rapid evolution with time-resolved spectroscopy from the Global Jet Watch observatories. Its H-alpha complex exhibited qualitatively different spectral signatures in the following weeks and months. We find that these H-alpha complexes are well described by the same five Gaussian emission components throughout the six months following eruption.  These five components appear to constitute two pairs of lines, from jet outflows and an accretion disc, together with an additional central component.  The correlated, symmetric patterns that these jet/accretion disc pairs exhibit suggest precession, probably in response to the large perturbation caused by the nova eruption. The jet and accretion disc signatures persist from the first ten days after brightening -- evidence that the accretion disc survived the disruption. We also compare another classical nova (V6568 Sgr) that erupted in July 2020 whose H-alpha complex can be described analogously, but with faster line-of-sight jet speeds exceeding 4000 km/s. We suggest that classical novae with higher mass white dwarfs bridge the gap between recurrent novae and classical novae such as YZ Reticuli.
\end{abstract}




\section{Introduction}
When \cne\ erupt, their optical luminosities increase dramatically and rapid spectral changes are observed which are exemplified in their evolving Balmer \halpha\ complexes. Changes occur on rapid and varied timescales ranging from hours to weeks. One such \cn, \mgab, underwent an outburst in July 2020 and we implemented a sustained, time-resolved, spectroscopic follow-up programme with the \GJW\ observatories, monitoring it extensively throughout the first few months after it was discovered. This revealed properties of key dynamical components that drive its observed evolution.  \mgab\ constitutes a particularly convenient example of a very energetic classical nova, having broad emission lines (FWHM $\sim$ \kms{3000}). The high speeds in this \cn\ make it possible to discern distinctly resolved components: we reduce the dimensionality of the data by fitting composite models of multiple \gaussians\ to the \halpha\ profile.  This reveals that much of the dramatic spectral behaviour can be explained by simple changes to these underlying components giving insights into the underlying dynamical story.

In Section \ref{sec:observations}, we outline our instrumentation and observations. We present our methodology in Section \ref{sec:fitting_methods} and the results of our fits in Section \ref{sec:fitting_results}. In Section \ref{sec:physical_interpretation} we suggest a physical interpretation for \mgab\ comprising a prominent and precessing accretion disc, and jets, in the aftermath of the nova event. We confront this model for \mgab\ with data for another \cn, PNV J17580848-300537, which also erupted in July 2020, in Section \ref{sec:pnvj_confirmation}. We summarise our conclusions in Section \ref{sec:conclusions}.

\subsection{Ejecta of \cne} \label{sec:spherical_shell}
It is widely agreed that many \cn\ eruptions eject roughly spherical shells, which move at slow speeds of up to (\about{\kms{1000}}). The evidence includes remnant shells of past novae \citep{Krautter2002,Santamaria2020}, and  absorption lines in the early spectra around maximum and immediately following it. These absorption lines often show slow and fast components, known as the principal and diffuse systems \citep{Mclaughlin1942,Williams1992}, and are linked to a slow, equatorial outflow and a faster polar outflow respectively. In some novae, these flows are thought to collide \citep{Williams2010}, causing shocks. Recent multiwavelength studies have shown correlations between gamma-ray detections interpreted as shocks between the various ejecta, and changes in the light curve \citep{Aydi2020a}. \citet{Slavin1995} present evidence that in addition to ellipsoidal shells, long tails of emission can appear. A popular idea is that in addition to an initially approximately spherical slowly-ejected shell, there is a significantly faster outflow along the polar axes which can disrupt the shell; this picture is clearly supported in the case of \DQHer\ in figure 3 of \citet{Toala2020}.

\subsection{Jets and accretion discs in novae}
The launch of jets in astrophysical systems undergoing accretion is widely accepted. However, typical cataclysmic variable systems have been thought to be accreting at too low a rate to launch jets according to established disc theory. \Cne\ could therefore present an important window on the emergence and survival of the jet phenomenon, because they have recently undergone disruptive eruption and associated enhanced accretion, which under favourable circumstances may put these systems over the threshold for the jet-launching mechanism to be triggered \citep{Lasota2005}.

There have been theoretical predictions and suggested observations of jets in \cne: \citet{Retter2004} discusses the theoretical viability of jets, using the V1494 Aql eruption in 1999 \citep{Iijima2003} as an example. \citet{Csak2005} find that there is evidence in favour of a link between the transition phase in novae and the establishing of an accretion disc. \citet{Kawabata2006} show jet-like winds in V475 Scuti, a moderately fast but otherwise unremarkable \cn\ that erupted in 2003. They used Okayama Astrophysical Observatory spectropolarimetry to show that the nova had blue and red jets with a constant position angle of linear polarisation.

In \novasgr, \citet{Harvey2018} presents a jet-like biconical outflow as an explanation for the profiles of [\ion{O}{iii}], although these model-dependent fits do not capture the asymmetry fully (see their figures 7 \& 8). In the following paper in this series (McLoughlin et al, in prep), we present \novasgr\ in the context of the model presented in this paper. We also note \citet{Ribiero2013} present modelling of a fast bipolar outflow based on their fits to emission line structures.

There is also evidence of  jets in \rne. \citet{Sokoloski2008} present direct evidence of highly collimated outflows and lobes in radio images of \rn\ \RSoph, having milli-arcsecond resolution, captured by the \vlba. These lobes were found to be moving across the sky with proper motions implying  underlying jet launch speeds of a few thousand \kms{}. Figure 5 of \citet{Taylor1989} shows a radio map of the object, with clearly aspherical lobes apparent within the first few months of the 1985 outburst. \citet{Rupen2008} observe synchrotron jets with the \vlba, noting that this implies emission is occurring far from the central source. This has important consequences, as it implies that fast jets may extend beyond the inner system long before the photosphere fully recedes. \citet{Skopal2008} presents spectra of the \halpha\ complex of \RSoph, and their figure 3 showing this is qualitatively similar to what we see here in \mgab, although they do not present any fitting analysis or decomposition of the complex.

\citet{Darnley2017} discuss the jets in the \rn\ M31N 2008-12a, which is near the Chandrasekhar limit, and thus accretes at a high rate, hence enabling the very rapidly (yearly or faster) recurring eruptions. They found broad shoulders around \halpha\ (offset from line centre by about \kms{4800} and about \kms{5900}) and attributed these to fast bipolar outflow - intriguingly faster than the jets we might expect from a lower mass white dwarf such as those in \cn\ systems.

V1721 Aquilae, a very fast \cn, within three days of discovery, was found to host an accretion disc that was either re-established fast, or was indeed never fully disrupted \citep{Hounsell2011}. It seems probable then, that there exist generalised jet-like outflows in erupting nova systems, with faster and more collimated jets occurring in recurrent systems containing more massive white dwarfs, through very fast novae such as V1721 Aquilae with moderate jets, to more typical \cne, such as the one for which we present evidence in this paper.

\subsection{\Cn\ \mgab} \label{sec:mgab_info}
\mgab\ was discovered by Robert H. McNaught (Coonabarabran, NSW, Australia) at magnitude 5.3 on 2020 July 15.590 UT (CBET 4811). It is also identified as Nova Reticuli 2020, and MGAB-V207, with Right Ascension of 03 58 29.55 and Declination $-$54 46 41.2 (J2000). There is a light curve from the All Sky Automated Survey for SuperNovae (hereafter \asassn) for this object which first shows evidence of brightening at 2020-07-08.1708099, or JD 2459038.67081, which is the epoch we take as the zero point. For the rest of this paper, we use the convention that +10.3d should be interpreted as 10.3 days after this zero point. There is a four-day gap in \asassn\ observations (magnitude 15 at 2020-07-04.18, magnitude 6 at 2020-07-08.17), so it cannot be precluded that it could have started brightening as early as 2020-07-04.18 \citep{ATEL13867}.

\begin{figure*}
	\includegraphics[width=2\columnwidth]{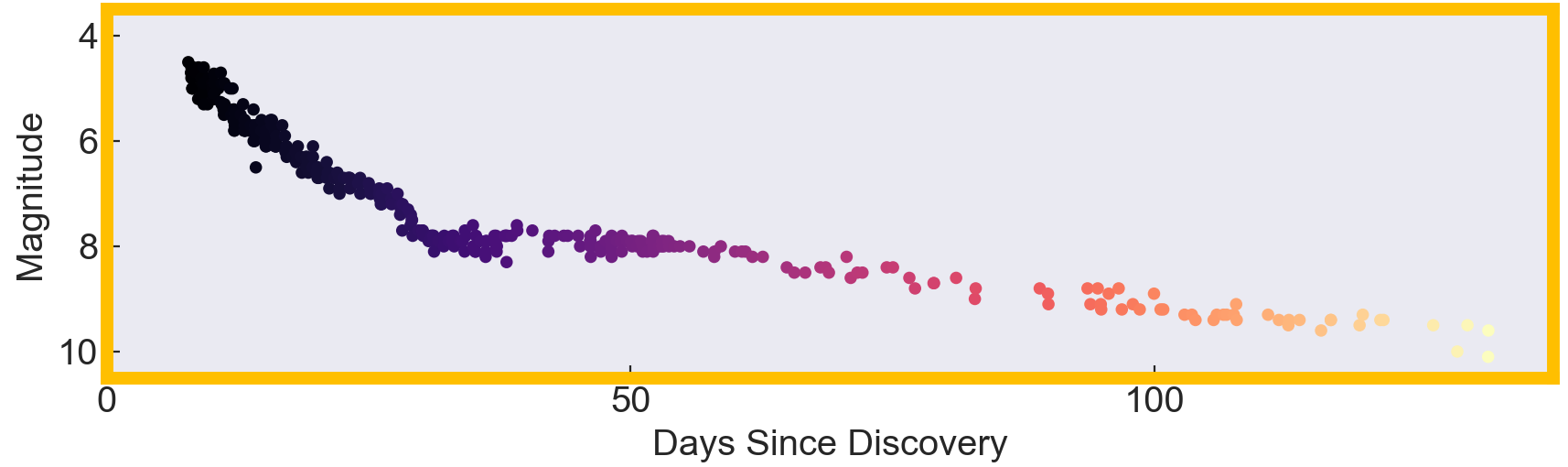}
    \caption{The AAVSO light curve of \mgab\, ranging from magnitude 4.5 to 10.1 in the visible band over the first 125 days since discovery.}
    \label{fig:mgab_light_curve}
\end{figure*}

Figure \ref{fig:mgab_light_curve} shows the AAVSO light curve\footnote{\href{https://aavso.org}{aavso.org}} of \mgab\ for the 125 days following eruption. The decline time of this \cn\ is $t_2 = 15$\,days. \mgab\ exhibits a plateau-type light curve \citep{Strope2010} between \about{\days{30}} and \about{\days{60}}, which may relate to the plateau in the light curve of \cn\ V407 Cyg, interpreted by \citet{Hachisu2012} as a surviving accretion disc emerging out of the receding photosphere.

This \cn\ was detected in gamma-rays between 2020-07-10 and 2020-07-15 by \fermilat\ \citep{ATEL13868}. \citep{ATEL13900} detected it in X-rays using \nustar, and found evidence for non-solar abundances of \ion{N}{}, \ion{O}{} and/or \ion{Fe}{}, in $67\,\si{\kilo\second}$ of total exposure commencing 2020-07-17.98. A light curve is available from \asassn\footnote{\href{https://asas-sn.osu.edu/light\_curves/ee130388-139d-4f3a-ab8a-7427a7545e21}{https://asas-sn.osu.edu/light\_curves/ee130388-139d-4f3a-ab8a-7427a7545e21}} for the 2000 days preceding eruption. We did not detect periodicities in the nova's quiescent luminosity in that dataset using the CLEAN algorithm together with deconvolution following \citet{Roberts1987}, but note that it remains at around 16 magnitudes. 

\citet{ATEL14067} notes that the X-ray luminosity of \mgab\ is significantly lower than expected, which may hint at the existence of a spatially extended accretion disc which could be obscuring a significant portion of the hot surface of the white dwarf. Coronal line emission is reported by \citet{ATEL14205}, who also comment on the absence of dust and on the appearance of double-peaked profiles with deep troughs at the centre that we explore in Section \ref{sec:central_component}. This X-ray phase, deemed by \citet{ATEL14043} to be the super-soft phase, commenced at \days{58}, indicating the epoch by which the photosphere has receded enough to see hydrogen burning on the surface of the white dwarf.

This \cn\ is in the GAIA DR2 catalogue \citep{GAIA_mission,GAIA_DR2} (source identification number: \#4731746232846281344), with a recorded parallax of $[0.3781 \pm 0.038]\,$mas, giving a distance of $2.7\,^{+0.4}_{-0.3}\,\si{\kilo\parsec}$ \citep{Bailer-Jones2018}. As it was not previously classified as a variable star, it does not have a GAIA light curve.

\subsection{\Cn\ \pnvj}
\Cn\ \pnvj, also known as Nova Sagittarii 2020 \#3, or PNV J17580848-300537, was discovered at magnitude 9.9 by Shigehisa Fujikawa on 2020 July 16.51939 UT (JD 2459047.01939). \pnvj\ has Right Ascension 17 58 08.48 and Declination $-$30 05 37.6 (J2000). It was classified as a \cn\ by \citet{ATEL13872} using high-resolution optical spectroscopy from the Southern African Large Telescope (SALT).

\section{Observations} \label{sec:observations}
This paper focuses on a sequence of observations of \mgab\ from \days{8.29} to the date of submission predominantly from the Aquila spectrographs in the Global Jet Watch observatories but supplemented by some \Hires\ spectra at the Rainbow Observatory (IAU \#430). A discussion of the comparison between the data from the different instruments is provided in Section \ref{sec:robustness}. The cumulative exposure time on this target thus far is 3.1\,weeks, across a range of different exposure times which lengthened as the \cn\ declined in brightness. Some spectra were intentionally saturated in \halpha\ so as to enhance the signal-to-noise in the less intense parts of the spectrum, and these spectra are used in the analysis of weaker lines.
 
\subsection{Global Jet Watch observatories}
The \GJW\footnote{\href{www.GlobalJetWatch.net}{www.GlobalJetWatch.net}} observatories are a network of five telescopes distributed in longitude (South Africa [GJW-SA], Chile [GJW- CL], east Australia [GJW-OZ], Western Australia [GJW-WA], India [GJW-IN]), described in Blundell et al (in prep). This configuration allows for sustained observations of an astrophysical target while the Earth rotates. Each observatory contains a Ritchey-Cretien carbon fibre telescope having a 0.5m-diameter primary mirror, and a fibre-fed spectrograph known as \Aquila\ (Lee et al, in prep). 

\subsection{Aquila spectroscopy}\label{sec:aquila}
The predominant data product used in this paper is a set of 5129 optical spectra captured by the \Aquila\ spectrographs. Our \Aquila\ observations start soon after discovery and last until the present day. The \Aquila\ spectra have a spectral resolution of $R \sim 4000$, and span the wavelength range from \angstrom{4930} to \angstrom{8840}. The camera in each Aquila spectrograph is either a FLI ML16200 or a FLI ML8300, and each is cooled to $-30^{\circ}\,$C by a Peltier cooler.

\subsection{LHires spectroscopy}\label{sec:lhires}
The Shelyak LHires instrument is a slit spectrograph, which uses an \ion{Ne}{} lamp for wavelength calibration arcs. It uses a ZWO ASI camera (model ASI294MM) cooled to $-15^{\circ}\,$C. While the slit width is selectable between several preset widths, it was set to $23\,\si{\micro\metre}$ for these observations. This instrument was commissioned to support the \Aquila\ spectrographs, complementing them with high-resolution ($R \sim 20,000$) auxiliary spectra. These spectra allow us to achieve a high level of confidence in our fits that deconstruct the highly time-sampled \Aquila\ spectra presented in Sections \ref{sec:fitting_methods} and \ref{sec:fitting_results}. It has been a fruitful strategy to obtain a high temporal density sustained dataset using \Aquila, with occasional supplementation from \Hires.   

\subsection{Epochs of observations}
\begin{figure}
	\includegraphics[width=\columnwidth]{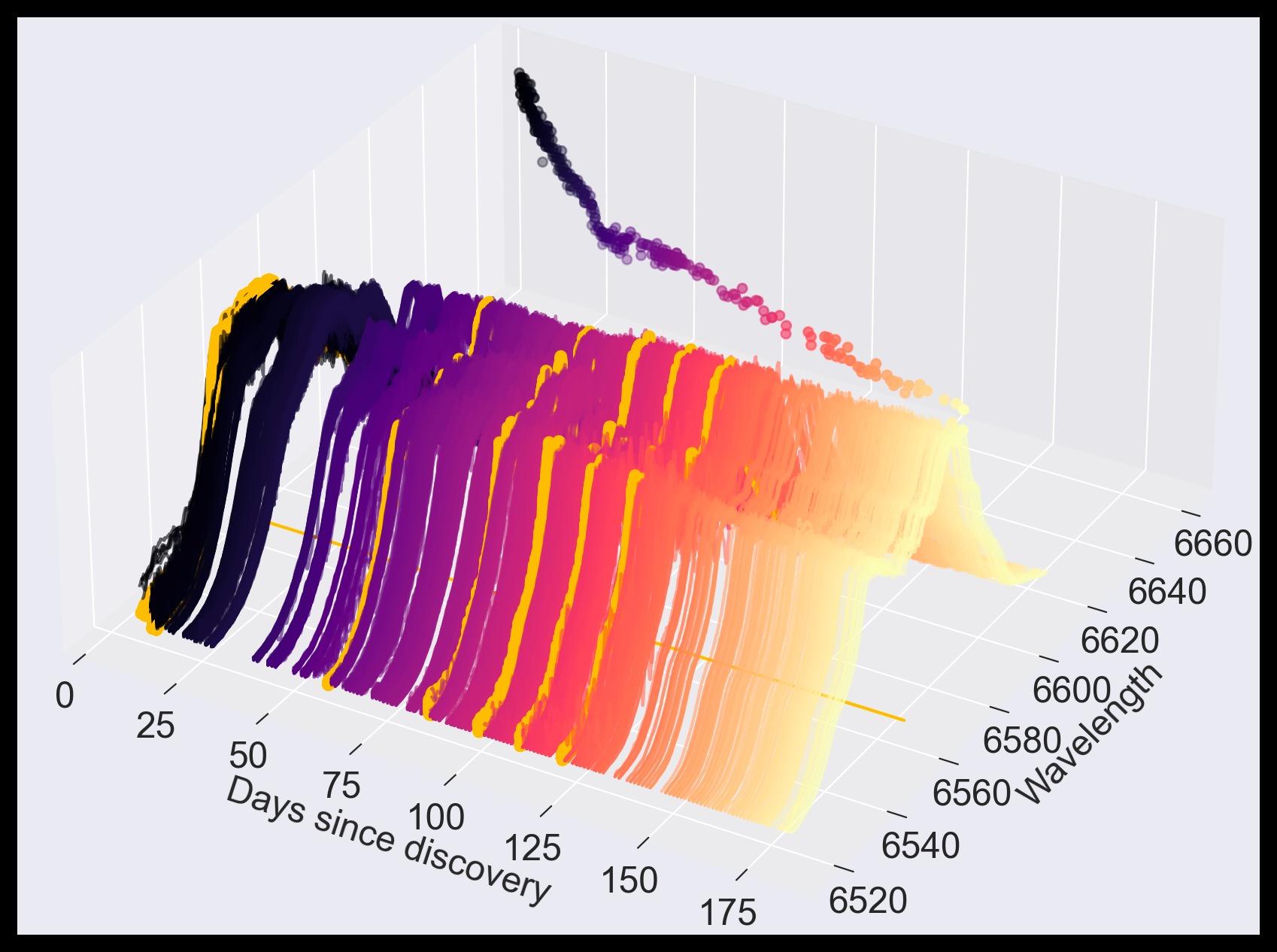}
    \caption{This plot indicates the epochs at which we have spectra. The \Hires\ spectra are depicted in gold, while the light curve and the \Aquila\ epochs are coloured according to their Julian Date. The rest wavelength of \halpha\ is projected to the floor as a fiducial marker. We reproduce for context the AAVSO light curve of \mgab\ against the back wall, coloured as in Figure \ref{fig:mgab_light_curve}}
    \label{fig:MGAB_observations}
\end{figure}

We present the epochs at which we have data in Figure \ref{fig:MGAB_observations}, marked against the AAVSO light curve on the back wall. The colour of the light curve at the back is mapped to Julian Day using the same colour map as that of the \Aquila\ observations themselves, shown as spectra in the foreground; this light curve is shown in Figure 1. The \Hires\ spectra are also shown, interspersed in gold. This gives an general impression of our dataset, as a continuous follow-up programme with highly regular observations and indicates how such an extended yet thorough campaign is made possible by the network of \GJW\ observatories, as there is typically at least one location with both local night-time and favourable observing conditions.

\subsection{Data reduction} \label{sec:data_pipeline}
The spectra were reduced with a bespoke data reduction pipeline called \codeendeavour, a full end-to-end image-processing solution which takes data captured at the telescopes, performs the reduction, and delivers 1D spectra as a product to the fitting applications. It is a bespoke pipeline suited to our specific instrumentation, but built on widely-used open-source scientific libraries available for the Python programming language. The packages used include NumPy \citep{numpy}, SciPy \citep{scipy}, Pandas \citep{pandas2010}, scikit-learn \citep{sklearn} and Astropy \citep{astropy2013,astropy2018}. The pipeline ensures every spectrum is dark and bias corrected, and calibrates wavelengths against Thorium-Krypton bulb arcs. The spectra are heliocentrically corrected to the barycentre of the solar system.

\subsection{Post-processing application} \label{sec:post_processing}
Post-processing analysis is performed by \codepoirot, which takes as an input the reduced 1D spectra from \codeendeavour. In this paper, we focus on the \halpha\ complex, and so data are typically normalised to the peak of \halpha. In the fits presented in this paper, the background has been fit with a simple constant offset, valid as the background does not vary strongly over the \halpha\ region. This is performed at the point of fitting the physical model, and is allowed to freely vary. We find that the background is consistently responsible for $<1\,\%$ of the signal in our region of interest, and so we subtract it.

During the data-cleansing step of \codepoirot, it removes any spectra with saturation on the feature of interest, or adverse weather effects (determined by manual inspection). In the later stages of the eruption when changes are happening on longer timescales, we sum the spectra taken at a given observatory on a given night. When necessary, as the nova became fainter, to further improve signal-to-noise, we reject any spectra which are not the longest exposure taken on a particular night.

\section{Fitting methods} \label{sec:fitting_methods}
Deducing the underlying physical model from spectral complexes is notoriously difficult in astronomy, in part because of the intrinsic richness of the phenomena, and in part because of the trade-off between spectral resolution and signal-to-noise.  Our \halpha\ spectra of \cne\ are no exception, with several dynamical components evolving rapidly. There are different strategies for deconstructing the underlying components that comprise emission lines profiles e.g.\ \citet{Schmidt2019}. We preferred to experiment with a simple set of \gaussians. The merits of this approach are demonstrated elsewhere e.g.\ \citet{Blundell2008,Blundell2020}.

\subsection{Composite fits}
\begin{figure}
	\includegraphics[width=\columnwidth]{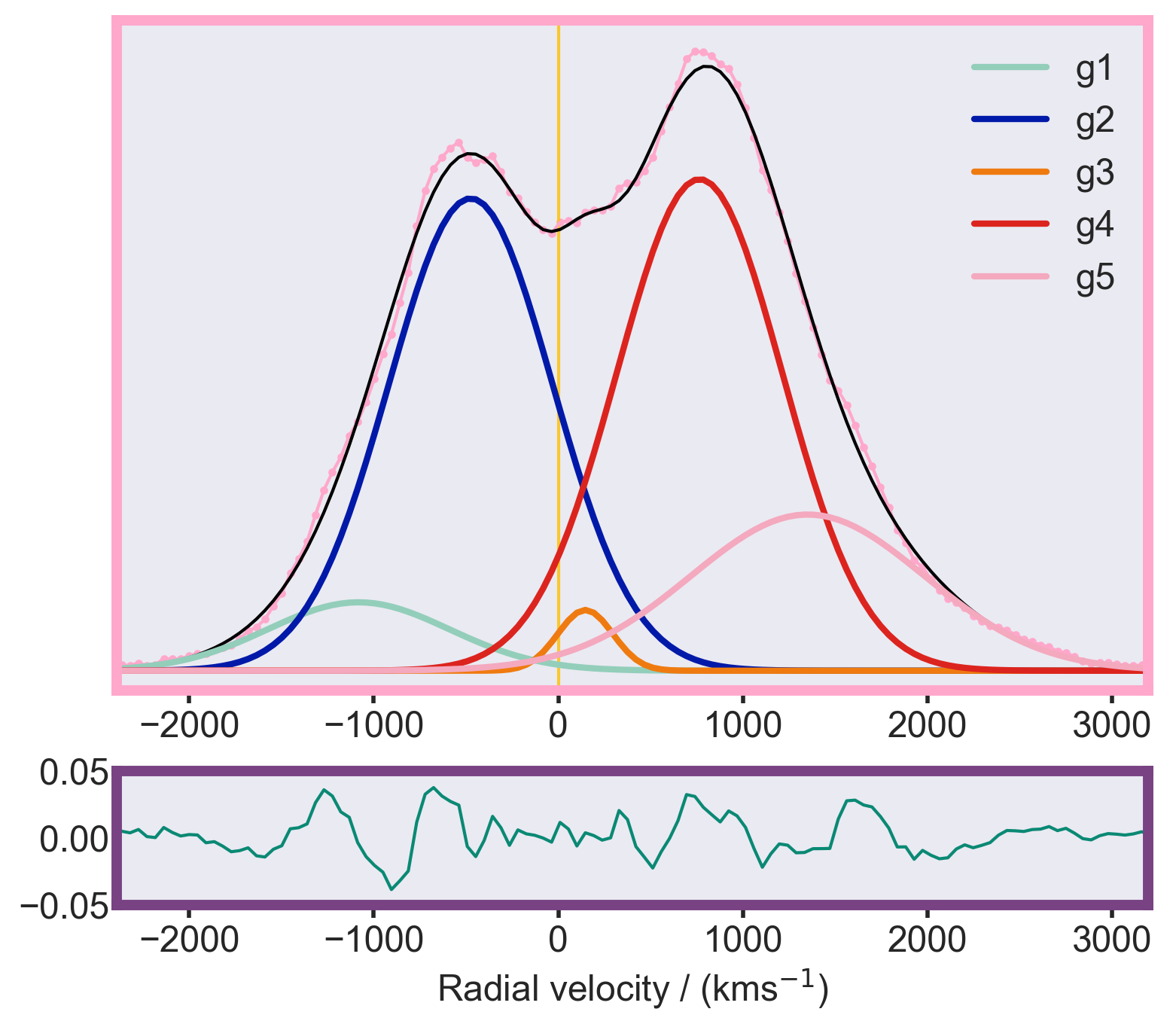}
    \caption{Fitted model showing two pairs of emission lines, and a central component. Spectrum (shown as pink points) taken at the GJW-OZ observatory, at \days{129.56}\ since discovery, with a 1000\,s exposure time.}
    \label{fig:MGAB_OZ_129_56_h_alpha_fit}
\end{figure}

\begin{figure}
	\includegraphics[width=\columnwidth]{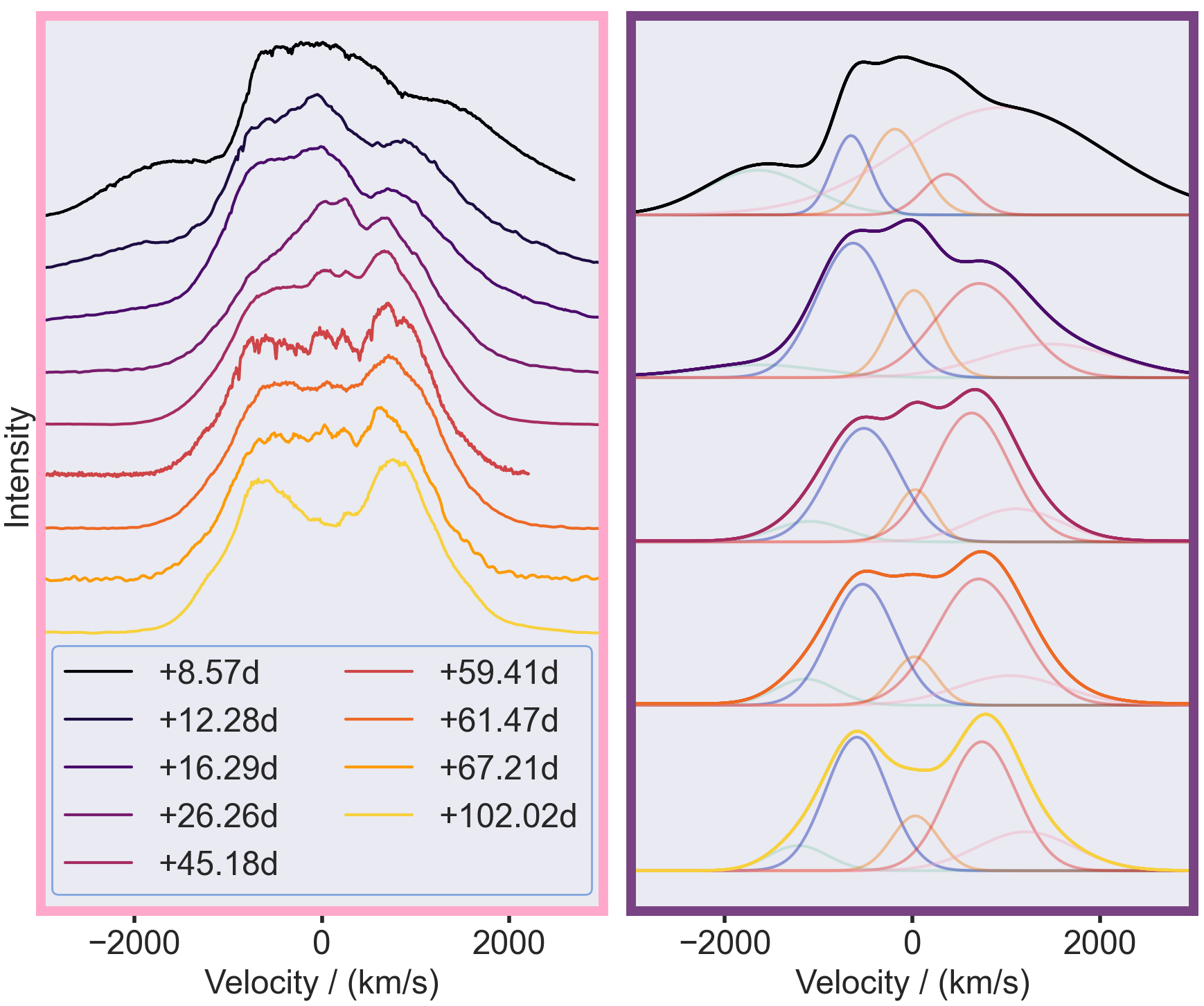}
    \caption{Sequence of H-alpha profiles. An example \Hires\ spectrum (fourth from the bottom) covers a shorter wavelength range than Aquila. These profiles can be deconstructed into a similar set of five emission \gaussian\ components. Section \ref{sec:fitting_results} presents these components in more detail.}
    \label{fig:mgab_sequence_halpha}
\end{figure}

We used multiple one-dimensional \gaussian\ models (and a subsequently subtracted constant offset background, discussed in Section \ref{sec:post_processing}) in a composite fit to our data, using the Levenberg-Marquardt algorithm with a least squares statistic \citep{astropy2018} to minimise the difference between observation and model. Residuals were analysed, and inspected to ensure that there only remained variation on velocity scales smaller than or comparable to the limit from spectral resolution.  See Section \ref{sec:robustness} for a discussion of the robustness of the fits. Figure \ref{fig:MGAB_OZ_129_56_h_alpha_fit} demonstrates typical results of our fitting template, with residuals plotted in the panel below. Figure \ref{fig:mgab_sequence_halpha} illustrates how the qualitatively changing profile of \mgab\ is nonetheless well-modelled by these five Gaussian components.

\subsection{Importance of coverage}
The extended coverage spanning 180\,days together with the consistency of the succession of fits were key to our confidence in these fits. While there are many possible fits which might reasonably approximate any one individual spectrum, we were able to reject most \textit{a priori} possible fits by comparing across time ranges within our dataset, shown in Figure \ref{fig:MGAB_observations} alongside the light curve. 

In Section \ref{sec:mgab_evolution} we discuss various changes which occur in the evolution of the fits to \mgab\ \halpha\ profiles. We are confident of variation timescales that are significantly sub-night, which would have been missed with sparser sampling.

\subsection{Robustness} \label{sec:robustness}
\begin{figure*}
	\includegraphics[width=2\columnwidth]{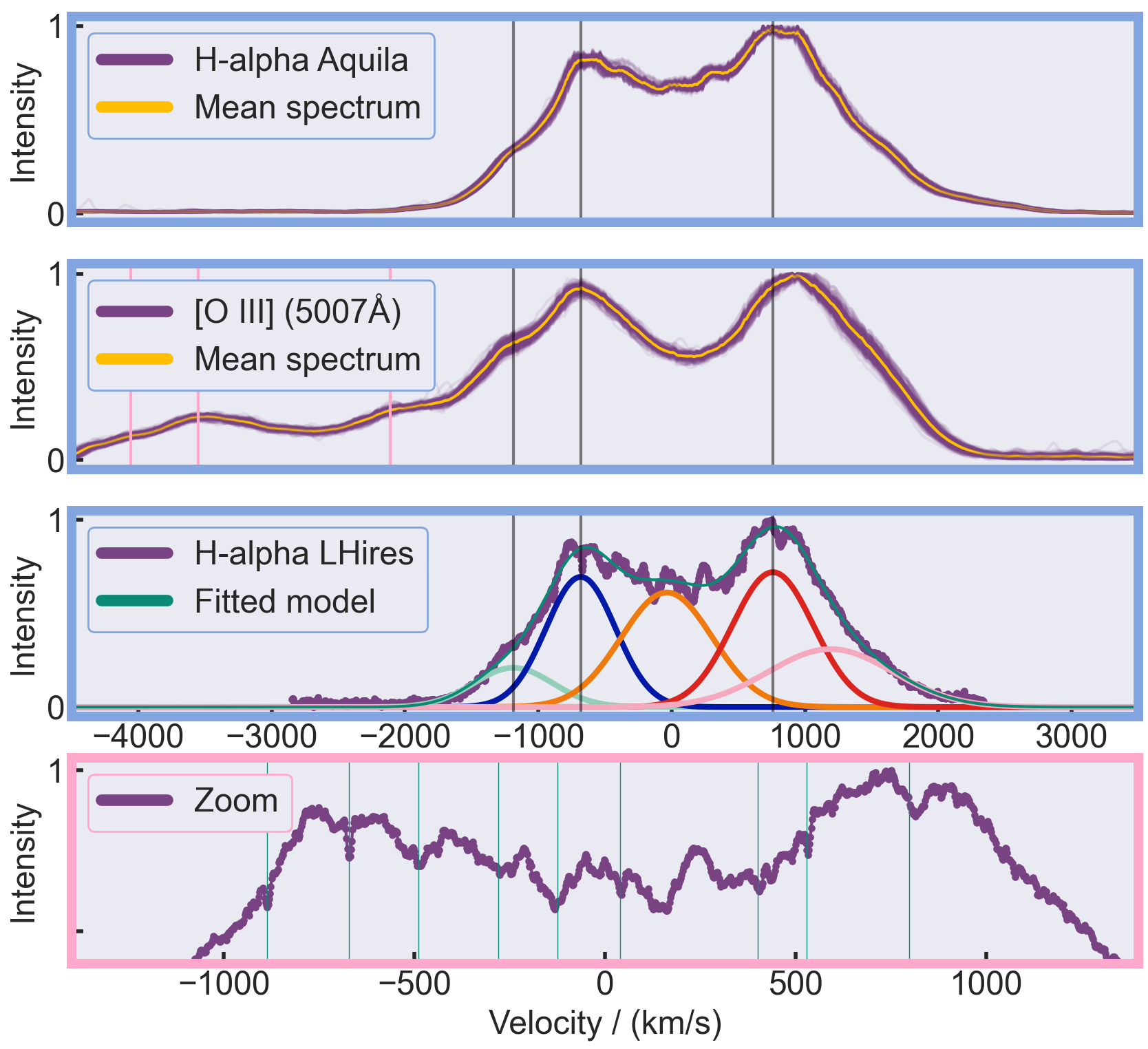}
    \caption{On the first three panels, we draw vertical lines marking peaks at radial velocity \kms{-1200}, \kms{-680} and \kms{760} for the relevant transition. Top panel: \halpha\ complex taken on an \Aquila\ instrument over the course of several nights (\days{120} to \days{160}) plotted in purple, with the mean of these overlaid in gold. Second panel: [\ion{O}{iii}] (\angstrom{4959} and \angstrom{5007}) complex in the same spectra, with corresponding sets of markers at the same radial velocities for each of these transitions. Third panel: an example \halpha\ spectrum from \Hires\ at roughly the same time but significantly higher resolution shown in purple, with the sum of five fitted \gaussian\ emission components overlaid in teal. The vertical markers match the centroids of three of these peaks. Fourth panel: a close-up of the top of the previous panel's \halpha\ complex, with known telluric lines denoted with vertical lines given by \citet{Curcioa}. Several of these have only slightly weaker doublet partners whose wavelengths are not known as accurately as the ones depicted, hence some of these lines appearing wider than others}. The close matching of the fit despite the telluric features confirms the appropriateness of our template model to explore the dynamics of this \cn. The wavelength alignment required no additional correction beyond the calibration as part of the \codeendeavour\ pipeline. Each spectrum is normalised to the peak of the complex being plotted, but the oxygen \angstrom{5007} peak is a factor of three weaker than that of \halpha.
    \label{fig:quality_control}
\end{figure*}

We now consider different signals that might give a minor influence to our determinations of the underlying \halpha\ emission components. Figure \ref{fig:quality_control} shows Earth's own atmospheric lines (bottom panel), which consist of many narrow absorption lines. These lines do not have a significant impact on the fits, as the second panel shows clearly. Many of these lines are in doublets, although these are rarely fully resolved, even by the \Hires\ instrument, as they absorb against the relatively dim nova at later epochs.

Another feature usually observed in the early spectra of \cne is blue-shifted absorption. Denoted the low- and high-velocity components respectively, the hundreds of \kms{} and the thousands of \kms{} absorption systems are thought to be due to the complex ejecta \citep{Arai2016}. These are often significant and usually need to be fitted for early spectra. In the particular case of our observations however, by the time of our our first spectrum and thereafter, only emission components dominate; any such blue-shifted absorption components had already disappeared.

We also carefully consider the possible presence of forbidden [\ion{N}{ii}] lines at \angstrom{6548} and \angstrom{6584}. These lines could be very problematic for fitting the broad \halpha\ complexes in \cne\ such as \mgab\ or \pnvj\ as a full width at half maximum of \about{\kms{3000}} at late times is easily enough to encompass these wavelengths. However, the nitrogen lines would correspond to \halpha\ speeds of \kms{-685} and \kms{959}, and should exhibit similar dynamics to \halpha. Indeed, Figure \ref{fig:quality_control} shows that the profiles of two forbidden [\ion{O}{iii}] lines (\angstrom{4959}, \angstrom{5007}) in the spectrum are strikingly similar to \halpha\ (the [\ion{O}{iii}] emission at \angstrom{5007} is a factor of three less strong than \halpha\ at late times for our spectra).  If the nitrogen \angstrom{6548} and \angstrom{6584} lines were strong, we would expect to see evidence of a similarly spaced pattern --- but there is no such evidence, and nor is there any emission component at the rest wavelengths either. 

\section{Results} \label{sec:fitting_results}
When experimenting with fits to the \mgab\ \halpha\ spectra with multiple \gaussians, there was one solution set which persisted throughout the entire time series. While individual parameters changed over time as the nova progressed, two pairs of lines behaved in a correlated fashion, leading us to believe that these fits do indeed reflect an underlying physical reality rather than a mathematical degeneracy. The qualitative shape of the overall spectra seems to change dramatically over the course of our observations (see Figure \ref{fig:mgab_sequence_halpha} for examples), but we show that only small changes to the \gaussian\ components reasonably explain these variations in an intuitive way. While of course it is possible to model \cn\ emission line profiles with spatial models of the ejecta, we believe that the results emerging here show that a much simpler picture can explain a lot of the spectral behaviour without presuming any geometry in advance.

\subsection{Best-fitting model}
\begin{figure*}
	\includegraphics[width=2\columnwidth]{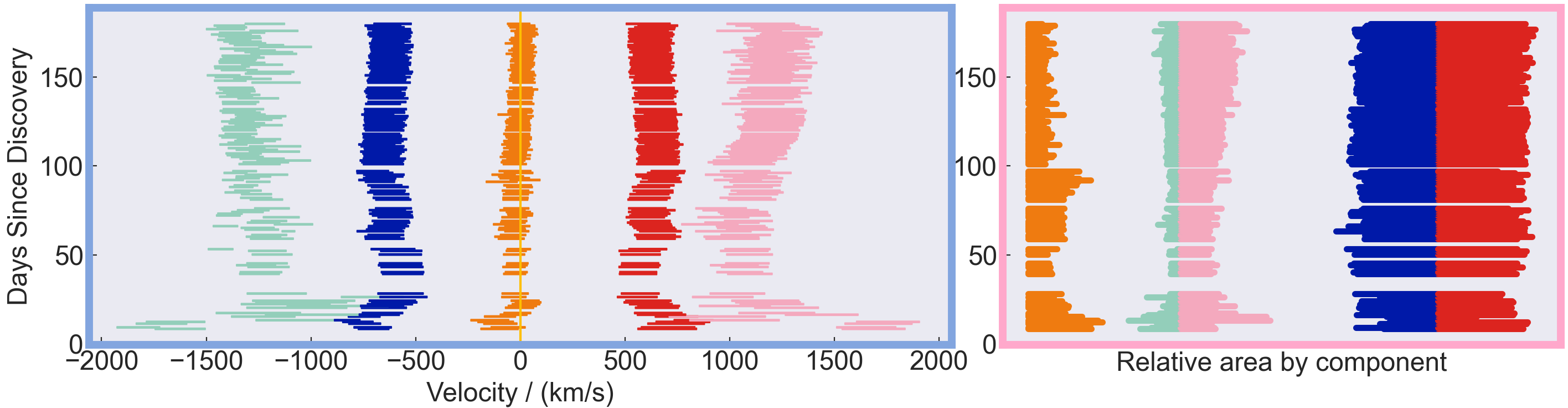}
    \caption{The left plot shows the centroid velocity of each of the five emission components relative to the centroid of the inner pair over time, ensuring that no subtle wavelength calibration trends are at play. The widths of the lines are proportional to the full width at half maximum of that component. The right plot shows how the area of each component develops with time, and the identified pairs of lines are plotted back-to-back to more easily see correlations. The width of each bar in this figure is proportional to the FWHM of the line, with an overall scaling that avoids horizontal overlap.}
    \label{fig:ideogram1}
\end{figure*}

The template which best fits our \halpha\ spectra consists of five \gaussian\ components in emission. In Figure \ref{fig:ideogram1}, the left plot is a convenient representation of the evolving characteristics of each fit because it represents the Doppler-shifting movements in the centroids of each of the five emission components, and reveals any time-varying patterns where these are present. The colour code for the five components is persistent throughout this paper: red and blue denoting the red- and blue-shifted double peaks of an inner pair of emission lines, and pink and turquoise correspondingly for an outer (higher-speed) pair. Orange is consistently the colour of the component roughly stationary with respect to the system, which we call the central component. There seems to be a natural pairing of the two outermost emission components, since the movement of their centroids is associated throughout time. The same is true for the inner pair (coloured red and blue). This leaves the single central component, close to the \halpha\ rest wavelength, which seems to vary only slightly in wavelength, though independently of the other four emission components. This is a persistent trait of the spectra throughout our spectroscopic monitoring.

What is truly remarkable about this five emission component model is that it fits the data well over a large range of dates, and for a variety of seemingly qualitatively different spectral complexes. The overall shape of the \halpha\ complex at early times displays broad emission, then from about \days{15} to \days{95} we observe a generally flat peak, with a surplus of red emission at about \kms{1000}. After \days{95} (long after the SSS switch-on) we see a sudden transition to a doubly-peaked profile with emission at \kms{-1000} and \kms{1000}. Our five peaks fit all of these rich data across the whole time series, with only minimal changes to the model parameters, which preserve the relationships between the pairs.

\begin{figure*}
	\includegraphics[width=2\columnwidth]{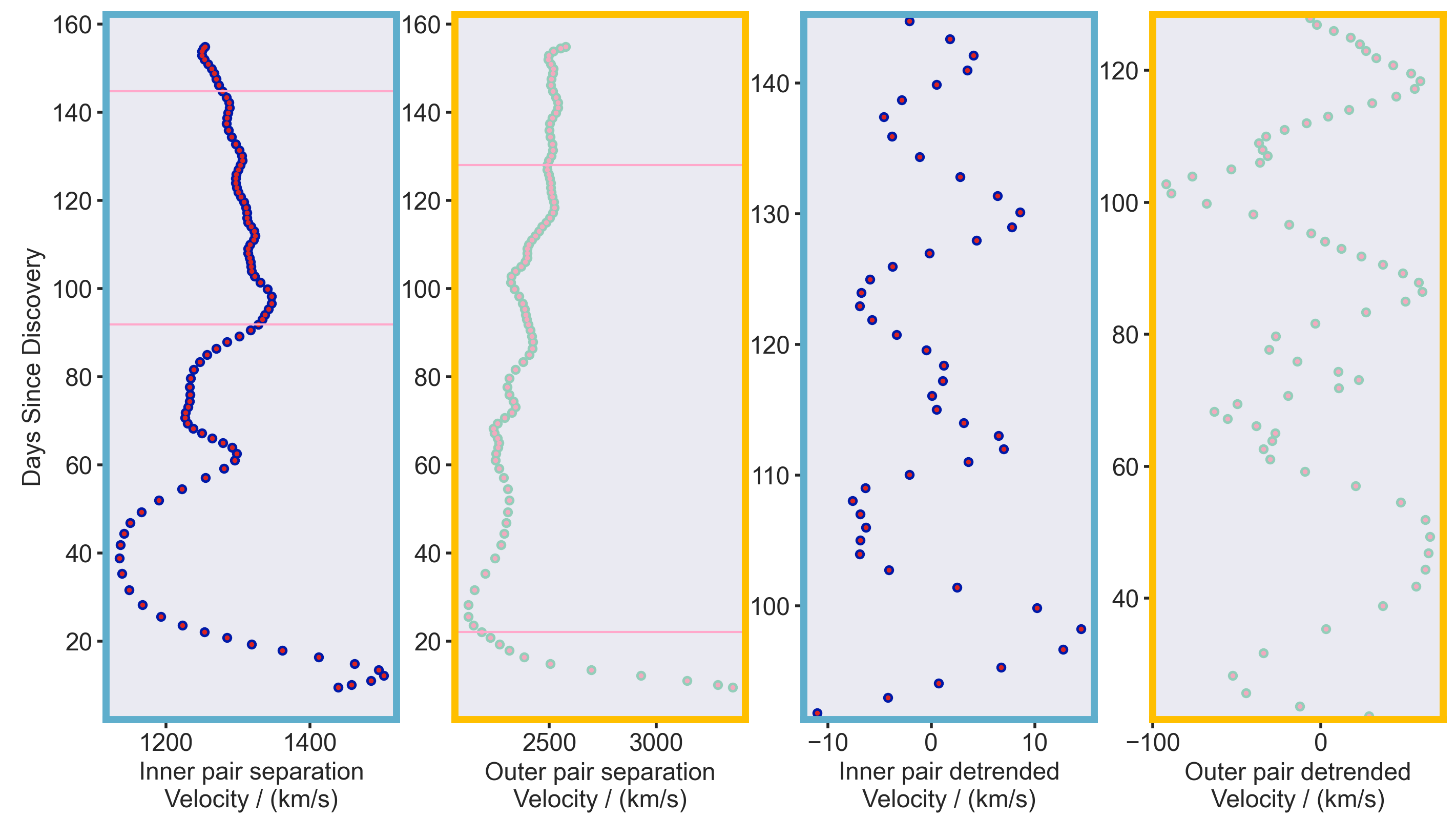}
    \caption{The two left panels show how the separations in velocity between each of the inner components and each of the outer components respectively vary over the full time range. These are smoothed using a mean filter with a moving window size of three observations. Each observation is the median of the highest-quality spectra we have for that observatory on each night. The right two plots represent the same underlying data as the left two plots, but over limited date ranges, and with the linear trend removed from each to show the oscillations more clearly. The date ranges used for the right-hand plots are represented by fiducial horizontal lines on their respective left-hand analogues.}
    \label{fig:ideogram2}
\end{figure*}

\subsection{Observed trends in derived parameters}\label{sec:mgab_evolution}
In the context of Figure \ref{fig:ideogram1} (left panel), we now consider not just how each emission component is behaving, but the relationships between them. We calculate relative metrics from the fitted parameters for each spectrum, and we plot these values over time in Figure \ref{fig:ideogram2}. This technique is essential for reducing the dimensionality of this highly complex dataset, and gives insights as to the underlying physics. The pertinent derived parameters include the differences between velocity centroids of the inner (blue, red) and outer (turquoise, pink) pairs of components, and the integrated area under each \gaussian\ component. The right panel of Figure \ref{fig:ideogram1} shows how the area of each \gaussian\ component varies with time.  The areas of the inner (blue, red) pair of components are remarkably steady relative to one another at every epoch. The central component, depicted in orange, shows a step-change decrease in area just before \days{100}. The redshifted outer component, depicted in pink, initially fluctuates to larger and then steadily smaller areas by \days{50}, after which it gradually increases in area.

Figure \ref{fig:ideogram2} depicts the time-variation of the difference in velocity between the red- and blue-shifted component centroids within each pair. We smooth this quantity (henceforth the separation) for each pair over time using a mean filter, and observe pronounced quasi-periodic oscillations in both the inner and outer pairs. Note that this is a separation in velocities rather than pure distances. These pairs behave in a similar fashion to each other, with large velocity separations at early times ($< $\,\days{50}) which reach a minimum separation within the first few weeks. They then both recover slightly, but continue to oscillate. The oscillations happen at different times for each pair, and the timescale of recurrence decreases over time, which we discuss further in Section \ref{sec:dynamical_changes}. 

At around \days{100} there is an abrupt change in behaviour, whereby the oscillation of the inner pair decreases in amplitude (moving to lower radial speeds) but the outer pair maintains its steady progression to higher-magnitude radial speeds. After this date, the predominant general trend is of much more steady increase in velocity by \about{\kms{2.5}} per day, but superimposed on this is the distinct but much more subtle oscillation of \about{\kms{10}} with a characteristic timescale of \about{10}\,days.

On removing the overall trends in the two leftmost panels of Figure \ref{fig:ideogram2} by subtracting a straight-line fit from each, we recover the underlying oscillations more clearly and these are plotted in the two rightmost panels for a restricted time range (marked in the left plots with horizontal lines). This representation shows the oscillations in the clearest manner. Upon folding by the strongest peaks in the Fourier transform, we could see that the period of both these pairs was in fact changing by several days, so this is in fact a quasi-periodic oscillation in the velocity separation. There were no statistically significant periodicities, most likely because the duration of each cycle was quite different from that of the preceding cycle. However, the pattern of irregular oscillations with similar velocity amplitude is clear, and we take this to mean that these pairings of lines reflect pairings in the underlying physical reality. In the two rightmost panels of Figure \ref{fig:ideogram2}, there also appear to be hints of smaller-scale nutations superimposed on the large oscillations, but we would need even finer time resolution to quantify these accurately and fully separate the various perturbations.

The inner pair velocity separation decreases by \kms{120} between \days{100} and \days{150}, while the outer pair separation steadily increases from \kms{2200} to \kms{2600} between \days{60} and \days{150}, which we discuss in more detail in Section \ref{sec:mgab_jets_model}.

\subsection{Dynamical changes} \label{sec:dynamical_changes}
What is fascinating about the oscillatory behaviour of the inner and outer pair seen in both Figure \ref{fig:ideogram1} and Figure \ref{fig:ideogram2} is that it is not fixed in either amplitude or period, but instead these quantities both decline over the course of the first six months. The implication here is that this reflects some property of the underlying system which is returning to an equilibrium. Furthermore, these timescales are substantially longer than both the canonical \cn\ binary orbit timescale (\about{4\,hr}) and the canonical white dwarf spin period in \cne\ (\about{76\,\si{\second}}), meaning this effect takes tens to hundreds of orbital periods to perform a cycle.

\section{Physical interpretation} \label{sec:physical_interpretation}
The data clearly fit well with a template of 5 \gaussian\ emission components: a central component, an inner pair of components with radial velocity relative to the centre of \about \kms{600}, and an outer pair of \about \kms{1200}.

There are several possible models which may initially appear to explain this clear signal. The key components in the system are thought to be the accretion disc, the hot white dwarf, the secondary star, and various forms of ejecta. We disregard the \halpha\ flux from to the secondary itself, since the magnitude of the whole system was $>=15$ during quiescence (\asassn\ data in Section \ref{sec:mgab_info}).

\subsection{Jets and accretion disc} \label{sec:mgab_jets_model}
The natural interpretation of the outermost widely-separated and broad emission lines is as oppositely-directed collimated outflows, which may reasonably be described as jets \citep{Iijima2003}. Given the prevalence of accretion discs within jet-ejecting systems, we posit that the inner pair of lines could arise from the accreting material either because of its rotation or winds off the accretion disc.

We propose that the five \gaussian\ emission components within the \halpha\ complex are best explained by a model involving jets, an accretion disc and an additional component henceforth the `jets' model. This model therefore posits that all five observed \gaussians\ have some connection with an accretion disc. The outer pair of lines (turquoise, pink) is caused by bipolar anti-parallel jets launched as a consequence of an accretion process. We therefore suggest that the inner pair of lines (blue, red in Figure \ref{fig:ideogram1}) may be a manifestation of the standard double-peaked rotating disc signature or the wind off this rotating matter.  This is a simple suggestion, and is exactly what is seen arising from the accretion disc in the microquasar \ssf\ \citep{Blundell2008}. As noted by \citet{Long2002}, the evidence for winds off the discs in cataclysmic variable systems is compelling. The fifth, central component has a significantly lower line-of-sight Doppler shift and may be a shell arising from local matter that is not dynamically important.

This model is supported by the fact that there seem to be distinctive epochs where characteristics change in a related way in each of the components, sometimes after a short delay; these changes occur most notably around \days{100}. This is explained naturally under the jets model, because the lines all have a common origin in the accretion disc. Furthermore, the evolution of the pairs of lines and their oscillatory behaviour makes most sense when considering them as being derived from a precessing disc.

After about \days{100}, we observe that the inner pair moves to steadily lower speeds, while the outer pair increases its velocity separation. 
\Cn\ events differ from normal cataclysmic variables by their enhanced mass-transfer rates. The increased accretion rate may lead to their exceeding the threshold necessary for launching jets \citep{Lasota2005}, a situation that can persist for at least a few months following eruption. Since not all the ejecta receive enough energy to be fully ejected, some gas can be trapped and forms an envelope surrounding the inner system. As this gas steadily rejoins the accretion disc, we expect that there exists a region of temporary overdensity, which would then feed the jet. What we propose is that the trends in pair speeds are caused by such an effect. The inner pair decreases in speed, as the accretion disc is depleted at its inner radius (where it is fastest). The outer pair increases in speed, as the accretion disc processes this extra material and launches increasingly powerful jets. This assumes the outer pair of lines are produced near the base of the jet, which is expected since at high rates of adiabatic expansion into relative vacuum, the plasma becomes optically thin.

Simulations conducted by \citet{Figueira2018} under particular conditions result in the accretion disc being fully disrupted by the ejecta within the first forty minutes after outburst. This occurs when the mass of the ejecta is assumed to be high compared with the mass of the accretion disc. A typical timescale for reforming a disrupted accretion disc has been estimated at around hundreds of years. However, Figueira et al mention, accretion discs have been observed in novae within months of eruption, implying that they either are re-established more rapidly than theoretical predictions or indeed are never fully disrupted. Moreover, it is possible that accretion can be efficacious even when the disc has not yet attained a state of perfect equilibrium. \citet{Hernanz2002} show the resumption of accretion after only a brief period of disruption. Also, the best-fitting model to the progenitor spectral energy distribution in \novacar\ includes an accretion disc, shown comprehensively in \citet{Wee2020} and used to explain changes in the colour evolution of that \cn\ around 50 days after outburst. The spectral signature of the accretion disc, or potentially the wind off its surface, is visible beyond the photosphere. The very early time-series data on \mgab\ shows a rapidly-evolving regime which may be caused by an accelerated resumption of accretion processes. 

A succession of over-pressured jet ejecta could plausibly expand into each other, causing shocks and hence particle acceleration giving rise to the X-ray shocks observed by \nustar\ reported in \citet{ATEL13900}. This is exactly analogous to the behaviour exhibited in the microquasar \ssf\ as its \halpha-emitting jet bolides collide into one another, causing non-thermal, polarised emission after shock acceleration \citep{Blundell2018}.

\subsection{Precession of the accretion disc and its jets} \label{sec:mgab_precession}
The leftmost teal panel in Figure \ref{fig:ideogram2} which plots the velocity separation of the inner pair with time shows oscillations on two different characteristic timescales, one being about 40\,days prior to about \days{100}, and thereafter about \days{10} (zoom-in in the third panel). These oscillations show a general tendency to decrease in both amplitude and period for the entire duration. We conjecture that the first phase of oscillations might be a response of the accretion disc to the rapid influx of material in the immediate aftermath of the eruption. The existence of the inner lines only a few days after the initial \cn\ event indicates a pre-existing (albeit perturbed) accretion disc in the \mgab\ system. After about \days{100}, the amplitude of the oscillations is significantly smaller, suggesting equilibration has happened to some extent.

The parameter space occupied by cataclysmic variables, and a discussion of the precessing accretion discs in these systems, is outlined in \citet{Patterson2001}. \citet{Charles2008} give a summary of the observational properties of accretion discs in X-ray binaries, many of whose time-varying/periodic characteristics have been interpreted as precession on a wide variety of different timescales. 

It is plausible that the oscillations seen in the leftmost teal panel in Figure \ref{fig:ideogram2} may arise from reactive precession in a (likely asymmetrically) perturbed accretion disc. It is therefore interesting to consider that the oscillations seen in the gold panels (2 and 4) of Figure \ref{fig:ideogram2} arise from a change in axis of the jet ejecta given the ongoing re-orientation of its launch platform i.e.\ accretion disc. In the microquasar \ssf, its accretion disc has been shown to exhibit a persistent periodic pattern of precession and nutation tightly coupled to that exhibited by its famous precessing jets (figure 2 of \citet{Blundell2008}). In contrast to the quasi-persistent periodic precession of \ssf, \mgab's accretion disc appears to precess on shortening timescales as a response to the disruption following the \cn\ outburst, so a tight link between this and consequent jet precession might not be expected. This can be especially hard to discern when the underlying period is itself decreasing (perhaps due to mass-loss through the jets themselves).  Regardless of the duration of the precession period, the accretion disc and jets are expected to exhibit a quarter-phase lag because of their perpendicularity from one another. It is therefore interesting to note in the leftmost two panels of Figure \ref{fig:ideogram2} that there are distinct hints of such a phase difference between jet oscillations and accretion disc oscillations that are consistent with such a picture.

\subsection{The behaviour of the central component} \label{sec:central_component}
One phenomenon we observe in the succession of \mgab\ spectra is that several emission species appear to transition from a profile with a central peak to one characterised by wider double-peaks. The first such transition event occurs suddenly at \days{63.21} in \halpha. It then recurs multiple times, most dramatically at \days{93.93} (lasting until \days{95.92}). At \days{96}, the transition repeats, steadily deepening across the next few hours and drastically deepening again at \days{101.9}. Eventually, the \halpha\ complex remains in a split state. This unsteady approach to eventual stability could be due to the temperature of the emitting body of gas dropping below a relatively sharp cutoff, with some fluctuations in temperature causing it to temporarily return above the threshold before settling below. 

\begin{figure*}
	\includegraphics[width=2\columnwidth]{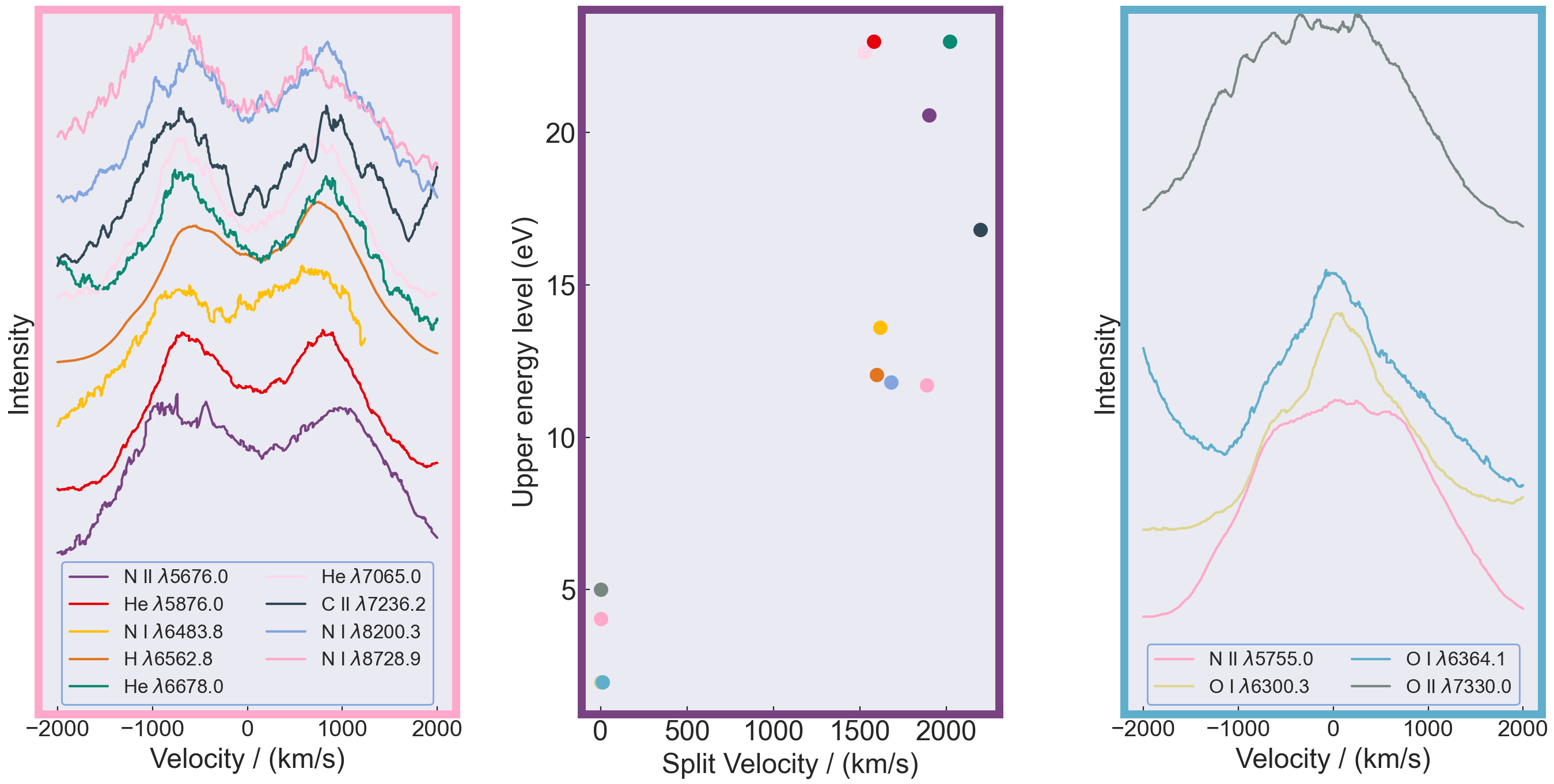}
    \caption{Average spectra of different emission species taken at \days{94} of \mgab. The left panel shows certain species which display the splitting phenomenon, the right panel shows the lines which do not exhibit this. The central panel shows the excitation energy of the upper level of the identified transitions responsible for each line, plotted against the velocity difference of the two peaks in the case of the split lines. Note that this is not to be confused with the quantum effect of line splitting, it is purely dynamical.}
    \label{fig:MGAB_94_32d_splitting}
\end{figure*}

Figure \ref{fig:MGAB_94_32d_splitting} shows the velocity profiles of the most prominent lines in our spectra. Some of the lines (shown in the left panel) display this split feature, and others do not (shown in the right panel). The central panel shows the split velocity (difference between the two peaks, zero if only one peak) against the upper energy level of the atomic transition responsible, giving a strong positive correlation. The transitions that do not appear to split show no qualitative sign of significant changes before and after \days{95}. 

By close examination of the fits to the \halpha\ complex in a time series around this epoch, we note that in fact the apparent splitting is caused by the central dynamical component growing weaker relative to the inner and outer pair. While the other lines are fainter and as such have lower signal to noise than \halpha\, they correspond very closely in velocity space, so we assume that they are also appearing to split because a central component is growing dimmer. 

Since the splitting seems to only occur in the higher upper energy level transitions, we believe that this is evidence that the origin of the central component has cooled to below $10\,$eV by \days{100}.

\subsection{Why are radio jets not observed in \cne?}
The existence of a jet whose radiant gas gives rise to \halpha\ emission does not necessarily imply the existence of a radio-emitting synchrotron jet.  In the case of the jets from the microquasar SS433, the former leads to the latter only because of the expansion of the jet bolides into one another causing shocks and consequent particle acceleration (Blundell et al 2018) evinced by particular signatures in the synchrotron polarisation coinciding with where the expansion and collision happens.  In the case of a binary whose \halpha\ emitting jet ejecta either (i) did not expand sufficiently as to collide into one another causing a shock-accelerated synchrotron emitting population or (ii) were not propagating through a magnetised region of space, no synchrotron radio jet would be observed.

We note that \citet{Toala2020} require a non-thermal component to fit the spectral shape of elongated X-ray emission of \DQHer\ shown in their figure 3; we conjecture that this X-ray emission arises from inverse Compton emission of spent synchrotron electrons.

\subsection{Why is the inner pair not just a spherical shell?} \label{sec:refute_spherical_ejecta}
Given the resolved images of approximately spherical shells around many \cne\ it might be tempting to associate the inner pair of emission lines with a spherically ejected shell discussed in Section \ref{sec:spherical_shell}. However, a ballistically-ejected outflow would not continue to modulate in an apparently symmetric way nor change speed several months after the eruption (Figure \ref{fig:ideogram2}, panel 3). Moreover, this could not explain the sometimes correlated changes in the characteristics of the inner pair and the outer pair of emission components. As described in Section \ref{sec:dynamical_changes}, the magnitude of oscillations in the relative velocity of the inner pair components consistently decreases over time. While this is expected behaviour for a precessing accretion disc re-establishing equilibrium, there is no clear explanation for why this would happen to opposite sides of a spatially-extended ejected shell.

In addition, a spherical shell would be too hot within the first few weeks of the eruption (\about{$10^8$}\,K) to emit \halpha, which requires a more modest temperature regime (a few $10^4$\,K). Indeed, the reason why we are able to see the jets in \halpha\ spectra is because they are far cooler, which is consistent with their having been launched by the accretion disc rather than being the products of thermonuclear runaway.

\subsection{Alternative models}
We considered several models before settling on the jets model. The primary such alternative model is that the inner pair is a slow-moving outflow, considered in Section \ref{sec:refute_spherical_ejecta}.

We also considered whether one of the pairs could be formed in a circumbinary disc. While we have seen preliminary evidence pointing towards circumbinary discs in other novae such as \novacar\ \citep{Mcloughlin2020}, the speeds involved here are significantly faster, implying either rotation within the binary system or high-speed outflow. 

Another alternative model is that one pair of lines is formed in an accretion disc while the other pair are the fast bipolar ballistic ejecta. The difference between this and our primary model is that in this case, the ejecta are not launched by the accretion, but are direct ejecta from the surface of the white dwarf. This model acknowledges that there has been a recent explosion and as such matter should be flying outwards, but it fails to explain why the inner pair and the outer pair should continue (i) at all and (ii) to oscillate, sometimes in related ways, to late times. Under this scenario, it would be difficult to explain why the bipolar outflow would be visible in \halpha, as it would be far too hot.

\section{Similar characteristics in \pnvj} \label{sec:pnvj_confirmation}
\begin{figure}
	\includegraphics[width=\columnwidth]{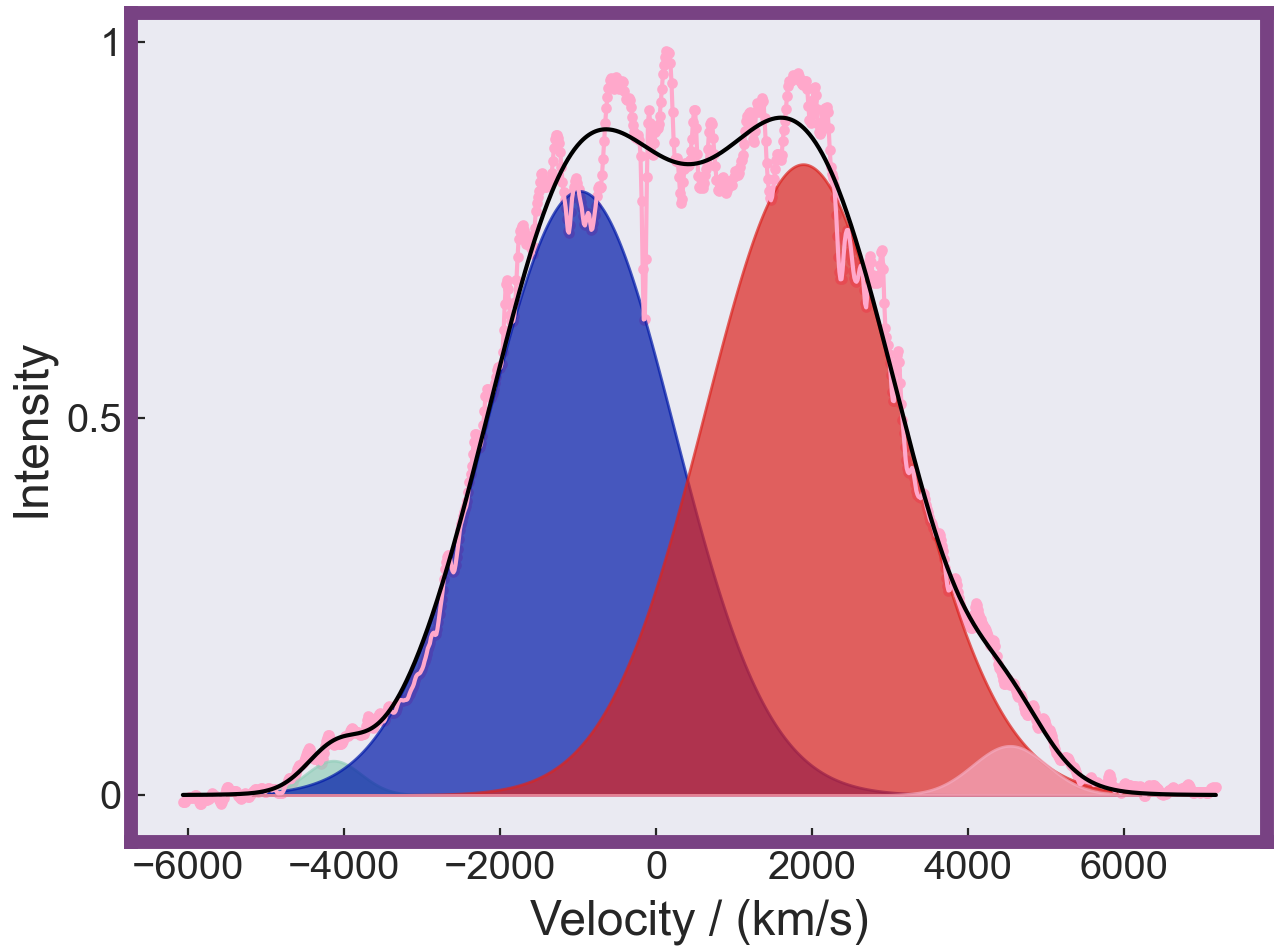}
    \caption{\pnvj\ spectrum taken at +1.03d after discovery. This fits the same model as \mgab, apart from the notable absence of a central component.}
    \label{fig:PNVJ_1_03d}
\end{figure}

Given the theoretical basis for jets in \cne\ set out in \citet{Kato2003}, it is tempting to consider whether this model applies in a prevalent manner, or if \mgab\ was an unusual event. In the same month, July 2020, there was another \cn, \pnvj, of the very fast variety. We confronted this with the same model, and discovered that while largely similar, it was, notably, missing the central component. The speeds of the outermost pair are \kms{-4100} and \kms{4500}, while the speeds of the inner pair are \kms{-1000} and \kms{1900}. Although \pnvj\ is substantially faster than \mgab, it nonetheless displays strikingly similar pairings. Figure \ref{fig:PNVJ_1_03d} shows that our model fits to this other \cn\ well, despite the vast difference in speed class. Note that the spectrum appears to be noisier than those of \mgab\ --- this is expected, because it is so much faster and because this was taken with the \Hires\ instrument which has a much smaller angstrom-to-pixel ratio. As such, there are fewer photons per wavelength bin, and the telluric absorption lines play a bigger role. Notwithstanding this, the same model fits, with the exception of a greatly diminished central component, which may be interpreted as any optically thick shell having disappeared. The two pairs of lines are very symmetric, and we could not obtain such a good fit without the outer pair; despite their relatively lower amplitude in comparison to the inner pair, they are nonetheless crucial. We conjecture that \pnvj\ bridges the gap in parameter space, specifically that of WD mass, between \mgab\ and the \rne.

\section{Conclusions} \label{sec:conclusions}
We have investigated the rapidly evolving spectral characteristics of the \cn\ \mgab, especially its \halpha\ complex, since shortly after its discovery in July 2020 with time-resolved spectroscopy from the \GJW.   We find that, remarkably, a model for the \halpha\ complex comprising the same family of \gaussians\ throughout the first six months fits the data well across a variety of qualitatively different spectral shapes.   These five components appear to be naturally categorised as two pairs of lines, representing an accretion disc and jet outflows, together with a small central component.

Correlated and symmetric changes in these two pairs of lines suggest that the accretion disc and jets appear to be precessing, probably in response to the recent large perturbation caused by the nova eruption.

The jet and accretion disc signatures are followed from the first spectra at around \days{10} until the present day, which we take as evidence that the accretion disc survived the blast, albeit in a non-equilibrium state for the first \about{100\,days}.

Confirmation of this interpretation for \mgab\ could be possible from future VLBI observations giving milli-arcsecond-scale resolution radio imaging akin to the images of the radio jets in \rn\ \RSoph\ observed by \citet{Sokoloski2008} but only if the \halpha\ jets give rise to particle acceleration and if there is a sufficiently strong magnetic field present.  However, it is undoubtedly fruitful to consider time-resolved spectroscopy from other \cne\ to establish the generality of \halpha\ jets, and we will present the results of such an analysis in a forthcoming paper.

\section*{Acknowledgements}
A great many organisations and individuals have contributed to the success of the Global Jet Watch observatories and these are listed on www.GlobalJetWatch.net but we particularly thank the University of Oxford and the Australian Astronomical Observatory. DM thanks the STFC for a doctoral studentship, and Oriel College, Oxford, for a graduate scholarship. We would also like to thank Robert H. McNaught for discovering this \cn\ during his bright transient search project.  We acknowledge with thanks the variable star observations from the AAVSO International Database contributed by observers worldwide and used in this research.

This work has made use of data from the European Space Agency (ESA) mission
{\it Gaia} (\url{https://www.cosmos.esa.int/gaia}), processed by the {\it Gaia}
Data Processing and Analysis Consortium (DPAC,
\url{https://www.cosmos.esa.int/web/gaia/dpac/consortium}). Funding for the DPAC
has been provided by national institutions, in particular the institutions
participating in the {\it Gaia} Multilateral Agreement. This work has made use of data from the All-Sky Automated Survey for Supernovae (ASAS-SN)\citep{Shappee2014,Kochanek2017}.

\section*{Data availability}
GAIA data publicly available at \href{https://gea.esac.esa.int/archive/}{https://gea.esac.esa.int/archive/}. ASAS-SN data publicly available at \href{https://asas-sn.osu.edu/light\_curves/ee130388-139d-4f3a-ab8a-7427a7545e21}{https://asas-sn.osu.edu/light\_curves/ee130388-139d-4f3a-ab8a-7427a7545e21}. AAVSO data publicly available at \href{https://aavso.org}{https://aavso.org}. The data underlying this article were provided by the \GJW\footnote{\href{www.GlobalJetWatch.net}{www.GlobalJetWatch.net}} by permission. Data will be shared on request to the corresponding author with permission of the \GJW. 




\bibliographystyle{mnras}
\bibliography{references2} 







\bsp	
\label{lastpage}
\end{document}